\documentclass[conference, twocolumn, 10pt]{IEEEtran} 
\IEEEoverridecommandlockouts

\usepackage[utf8]{inputenc}
\usepackage{amsmath}
\usepackage{amssymb}
\usepackage{url}
\usepackage{graphicx}
\usepackage{epstopdf}
\usepackage{caption}
\usepackage{comment}
\usepackage{subcaption}
\usepackage{multirow}
\usepackage{amssymb}
\usepackage{setspace}
\usepackage[noadjust]{cite}
\usepackage{xspace}
\usepackage{algorithm}
\usepackage{tabularx}
\usepackage{changepage}

\usepackage{graphicx,epsfig,bm}
\usepackage{cite}
\usepackage{amsmath,epsfig} 
\usepackage{times}
\usepackage{enumerate,type1cm}
\usepackage{amsfonts,relsize}
\usepackage{bm}
\usepackage{amssymb}
\usepackage{relsize}
\usepackage{fancybox}
\usepackage{algorithmic}
\usepackage{amssymb}
\usepackage{graphicx,epsfig,bm}
\usepackage{amsmath,epsfig} 
\usepackage{times}
\usepackage{enumerate,type1cm}
\usepackage{amsfonts,relsize}
\usepackage{bm}
\usepackage{amssymb}
\usepackage{relsize}
\usepackage{fancybox}
\usepackage{bbm}
\usepackage{tikz}
\usetikzlibrary{fit,positioning}
\usetikzlibrary{bayesnet}
\usetikzlibrary{shapes,decorations}
\usepackage[english]{babel}
\usepackage{xpatch}

\usepackage{pgf}
\usepackage{pgfplots}
\pgfplotsset{compat=1.18}
\usepackage{tikz}
\usepackage{subcaption}
\usepackage{eqparbox}
\usepackage{makecell}

\usepackage[xindy]{glossaries}
\usepackage{xcolor}

\def\BibTeX{{\rm B\kern-.05em{\sc i\kern-.025em b}\kern-.08em
		T\kern-.1667em\lower.7ex\hbox{E}\kern-.125emX}}
	
	\tikzset{
		startstop/.style={
			rectangle, 
			rounded corners,
			minimum width=3cm, 
			minimum height=0.5cm,
			align=center, 
			draw=black, 
		},
		process/.style={
			rectangle, 
			minimum width=3cm, 
			minimum height=0.5cm, 
			align=center, 
			draw=black, 
		},
		decision/.style={
			rectangle, 
			minimum width=3cm, 
			minimum height=0.5cm, align=center, 
			draw=black, 
		},
		arrow/.style={thick,->,>=stealth},
		dec/.style={
			ellipse, 
			align=center, 
			draw=black, 
		},
	}
	
	
	\makeatletter
	\def\BState{\State\hskip-\ALG@thistlm}
	\makeatother
	
	\DeclareCaptionLabelSeparator{periodspace}{.\quad}
		{\end{adjustwidth}}

\newcommand{\numAPs}{L}
\newcommand{\numAntennasPerAP}{N}
\newcommand{\numUEs}{K}
\newcommand{\numAntennasCPU}{M}

\newcommand{\tauu}{\tau_{\rm u}}

\newcommand{\chan}[1]{\mathbf{h}_{{#1}}}

\newcommand{\chanMat}[1]{\mathbf{H}_{{#1}}}

\newcommand{\chanMatCent}[1]{\mathbf{G}_{{#1}}}

\newcommand{\norm}[1]{\left\Vert#1\right\Vert}

\newcommand{\complexm}[2]{\mathbb{C}^{#1\times #2}}

\newcommand{\gram}[1]{#1^H#1}

\newcommand{\CN}[2]{\mathcal{CN}{(#1,#2)}}
\newcommand{\expect}[1]{\mathbb{E}\{#1\}}
\newcommand{\expectL}[1]{\mathbb{E}\left\{#1\right\}}
\newcommand{\trace}[1]{{\rm tr}\left(#1\right)}

\DeclareMathOperator*{\argmax}{arg\,max}

\newacronym{ai}{AI}{artificial intelligence}
\newacronym{aoa}{AOA}{angle-of-arrival}
\newacronym{aod}{AOD}{angle-of-departure}
\newacronym{ap}{AP}{access point}
\newacronym{asic}{ASIC}{application specific integrated circuit}
\newacronym{awgn}{AWGN}{additive white Gaussian noise}
\newacronym{blue}{BLUE}{best linear unbiased estimator}
\newacronym{ber}{BER}{bit error rate}
\newacronym{ce}{CE}{channel estimation}
\newacronym{cw}{CW}{continuous wave}
\newacronym{cdf}{CDF}{cumulative distribution function}
\newacronym{ced}{CED}{cumulative energy density}
\newacronym{clk}{CLK}{clock}
\newacronym{csi}{CSI}{channel state information}
\newacronym{csp}{CSP}{contact service point}
\newacronym{cpu}{CPU}{central processing unit}
\newacronym{crlb}{CRLB}{Cram\'er-Rao lower bound}
\newacronym{cu}{CU}{central unit}
\newacronym{dc}{DC}{direct current}
\newacronym{dl}{DL}{downlink}
\newacronym{dlt}{DLT}{Distributed Ledger Technology}
\newacronym{dp}{DP}{differencial privacy}
\newacronym{dram}{DRAM}{dynamic random-access memory}
\newacronym{du}{DU}{distributed unit}
\newacronym{ecc}{ECC}{elliptic curve cryptography}
\newacronym{ecdlp}{ECDLP}{elliptic curve discrete logarithm problem}
\newacronym{ecsp}{ECSP}{edge computing service point}
\newacronym{eirp}{EIRP}{equivalent isotropically radiated power}
\newacronym{em}{EM}{electromagnetic}
\newacronym{en}{EN}{energy neutral}
\newacronym{e2e}{E2E}{End-to-End}
\newacronym{fa}{FA}{federation anchor}
\newacronym{fdd}{FDD}{frequency division duplexing}
\newacronym{fhss}{FHSS}{frequency hopping spread spectrum}
\newacronym{fl}{FL}{federated learning}
\newacronym{gdpr}{GDPR}{general data protection regulation}
\newacronym{ia}{IA}{initial access}
\newacronym{ic}{IC}{integrated circuit}
\newacronym{ioe}{IoE}{internet of everything}
\newacronym{iot}{IoT}{internet of things}
\newacronym{id}{ID}{information decoding}
\newacronym{kpi}{KPI}{key performance indicator}
\newacronym{lca}{LCA}{life cycle assessment}
\newacronym{ls}{LS}{least squares}
\newacronym{lis}{LIS}{large intelligent surface}
\newacronym{liion}{Li-ion}{lithium-ion}
\newacronym{llr}{LLR}{log-likelihood ratio}
\newacronym{los}{LoS}{line-of-sight}
\newacronym{lmo}{LMO}{lithium ion manganese oxide}
\newacronym{map}{MAP}{maximum aposteriori}
\newacronym{mf}{MF}{matched filter}
\newacronym{ml}{ML}{machine learning}
\newacronym{mle}{MLE}{maximum likelihood estimator}
\newacronym{mvu}{MVU}{minimum variance unbiased}
\newacronym{mse}{MSE}{mean square error}
\newacronym{mmse}{MMSE}{minimum mean square error}
\newacronym{lmmse}{LMMSE}{linear minimum mean-square error}
\newacronym{ldpc}{LDPC}{low-density parity-check}
\newacronym{mmwave}{mmWave}{millimeter wave}
\newacronym{mimo}{MIMO}{multiple-input multiple-output}
\newacronym{miso}{MISO}{multiple-input single-output}
\newacronym{mpc}{MPC}{multipath component}
\newacronym{mrc}{MRC}{maximum ratio combining}
\newacronym{mrt}{MRT}{maximum ratio transmission}
\newacronym{nmse}{NMSE}{normalized mean square error}
\newacronym{ofdm}{OFDM}{orthogonal frequency-division multiplexing}
\newacronym{ota}{OTA}{over-the-air}
\newacronym{nb}{NB}{narrowband}
\newacronym{nr}{NR}{New Radio}
\newacronym{pa}{PA}{power amplifier}
\newacronym{pdp}{PDP}{power delay profile}
\newacronym{per}{PER}{packet error rate}
\newacronym{pet}{PET}{privacy enhancing technology}
\newacronym{pki}{PKI}{public key infrastucture}
\newacronym{pl}{PL}{path loss}
\newacronym{pc}{PC}{pilot count}
\newacronym{qrrls}{QR-RLS}{QR decomposition based recursive least squares}
\newacronym{qam}{QAM}{quadrature amplitude modulation}
\newacronym{re}{RE}{radio element}
\newacronym{rf}{RF}{radio frequency}
\newacronym{rfid}{RFID}{radio frequency identification}
\newacronym{ris}{RIS}{reflective intelligent surface}
\newacronym{rms}{RMS}{root mean square}
\newacronym{rls}{RLS}{recursive least squares}
\newacronym{rss}{RSS}{received signal strength}
\newacronym{rssi}{RSSI}{received signal strength indicator}
\newacronym{rw}{RW}{RadioWeaves}
\newacronym{sa}{SA}{synchronization anchor}
\newacronym{sdg}{SDG}{Sustainable Development Goal}
\newacronym{sdn}{SDN}{software-defined network}
\newacronym{se}{SE}{spectral efficiency}
\newacronym{ser}{SER}{symbol error rate}
\newacronym{sinr}{SINR}{signal-to-interference-plus-noise ratio}
\newacronym{siso}{SISO}{single-input single-output}
\newacronym{slam}{SLAM}{simultaneous localization and mapping}
\newacronym{snr}{SNR}{signal-to-noise ratio}
\newacronym{srls}{SRLS}{standard recursive least squares}
\newacronym{svd}{SVD}{singular value decomposition}
\newacronym{sp}{SP}{service point}
\newacronym{sram}{SRAM}{static random-access memory}
\newacronym{tdd}{TDD}{time division duplexing}
\newacronym{tdoa}{TDOA}{time-difference-of-arrival}
\newacronym{toa}{TOA}{time-of-arrival}
\newacronym{tx}{TX}{transmitter}
\newacronym{rx}{RX}{receiver}
\newacronym{ue}{UE}{user equipment}
\newacronym{uhf}{UHF}{ultra high frequency}
\newacronym{ula}{ULA}{uniform linear array}
\newacronym{upa}{UPA}{uniform planar array}
\newacronym{ul}{UL}{uplink}
\newacronym{ura}{URA}{uniform rectangular array}
\newacronym{uwb}{UWB}{ultrawideband}
\newacronym{wb}{WB}{wideband}
\newacronym{wpt}{WPT}{wireless power transfer}
\newacronym{zf}{ZF}{zero forcing}

\newacronym{lpt}{LPT}{laser power transfer}
\newacronym{rfpt}{RFPT}{radio frequency power transfer}
\newacronym{ipt}{IPT}{inductive power transfer}
\newacronym{cpt}{CPT}{capacitive power transfer}
\newacronym{apt}{APT}{acoustic power transfer}
\newacronym{cfo}{CFO}{carrier frequency offset}
\newacronym{lo}{LO}{local oscillator}
\newacronym{if}{IF}{intermediate frequency}
\newacronym{duc}{DUC}{digital up conversion}
\newacronym{ddc}{DDC}{digital down conversion}
\newacronym{dsp}{DSP}{digital signal processing}
\newacronym{vco}{VCO}{voltage-controlled oscillator}
\newacronym{pll}{PLL}{phase-locked loop}
\newacronym{lna}{LNA}{low noise amplifier}
\newacronym{dac}{DAC}{digital-to-analog converter}
\newacronym{adc}{ADC}{analog-to-digital converter}
\newacronym{de}{DE}{drain efficiency}
\newacronym{pae}{PAE}{power-added efficiency}
\newacronym{sar}{SAR}{specific absorption rate}
\newacronym{mppt}{MPPT}{maximum power point tracking}
\newacronym{isi}{ISI}{intersymbol interference}
\newacronym{pps}{PPS}{pulse per second}
\newacronym{icnirp}{ICNIRP}{International Commission on Non-Ionizing Radiation Protection}
\newacronym{fd}{FD}{Front-haul distance}
\newacronym{wd}{WD}{Wireless distance}
\newacronym{ntp}{NTP}{Network Time Protocol}
\newacronym{ptp}{PTP}{Precision Time Protocol}
\newacronym{wr}{WR}{White Rabbit}
\newacronym{wlan}{WLAN}{wireless local-area network}

\title{Resource Efficient Over-the-Air Fronthaul Signaling for Uplink Cell-Free Massive MIMO Systems}
\author{Zakir Hussain Shaik$^{*}$, Sai Subramanyam Thoota$^{*}$, Emil Bj\"{o}rnson$^{\dagger}$, and Erik G. Larsson$^{*}$\\
	$^{*}$Department of Electrical Engineering (ISY), Link{\"o}ping University, Link{\"o}ping, Sweden\\$^{\dagger}$Department of Computer Science, KTH Royal Institute of Technology, Stockholm, Sweden\\
	Emails: zakir.hussain.shaik@liu.se, sai.subramanyam.thoota@liu.se, emilbjo@kth.se, erik.g.larsson@liu.se\thanks{Zakir Hussain Shaik, Sai Subramanyam Thoota, and Erik G. Larsson were supported in part by KAW foundation and ELLIIT.}\thanks{Emil Bj\"{o}rnson was supported by the Grant 2019-05068 from the Swedish Research Council.}}
 \allowdisplaybreaks
\begin{document}

\maketitle
\date{}
\begin{abstract}

We propose a novel resource efficient analog \gls{ota} computation framework to address the demanding requirements of the \gls{ul} fronthaul between the \glspl{ap} and the \gls{cpu} in cell-free massive \gls{mimo} systems. We discuss the drawbacks of the wired and wireless fronthaul solutions, and show that our proposed mechanism is efficient and scalable as the number of \glspl{ap} increases. We present the transmit precoding and two-phase power assignment strategies at the \glspl{ap} to coherently combine the signals \gls{ota} in a spectrally efficient manner. We derive the statistics of the \glspl{ap}' locally available signals which enable us to to obtain the analytical expressions for the Bayesian and classical estimators of the \gls{ota} combined signals. We empirically evaluate the \gls{nmse}, \gls{ser}, and the coded \gls{ber} of our developed solution and benchmark against the state-of-the-art wired fronthaul based system.
\end{abstract}

\section{Introduction}
Cell-free massive \acrfull{mimo} systems is one of the technologies envisioned to combat the low \gls{sinr} seen by the cell-edge users in conventional multi-user \gls{mimo} systems. 
However, achieving the performance of fully centralized decoding with a reduced fronthaul communication overhead between the \glspl{ap} and the \gls{cpu} is crucial to realize its full potential~\cite{demir2021foundations,Ammar2022CFSurvey}. With high-speed wired fronthaul links, it is possible to communicate all the \gls{csi} and the received signals from the \glspl{ap} to the \gls{cpu} and perform the channel estimation and data detection. However, this is expensive and needs a large number of wired interconnects between the \glspl{ap} and the \gls{cpu}. Moreover, placing the \glspl{ap} to suit the wired fronthaul availability may be suboptimal for the system performance. Also, when the number of \glspl{ap} increases, which is imperative to meet the demands of future generation communication systems, a wired fronthaul with a predefined capacity is unscalable.



In this paper, we consider an uplink cell-free massive \gls{mimo} system where multiple multi-antenna \glspl{ap} serve multiple single-antenna \glspl{ue}. The \glspl{ap} communicate with the \gls{cpu} through wireless fronthaul links. 
To expose the main idea of this paper, we assume that each \gls{ap} has a separate \gls{rf} chain to communicate with the \gls{cpu} and uses a different frequency band for the fronthaul signaling. Our goal is to devise a spectrally efficient scheme to communicate the local information from the \glspl{ap} to the \gls{cpu} that is sufficient to decode the data. Our proposed \acrfull{ota} framework enables the \glspl{ap} to transfer their local received signals and the \gls{csi} to the \gls{cpu} using the same set of resources which results in huge savings in the communication and computation overhead.

Several solutions to handle the fronthaul related issues have been studied in the literature~\cite{Ammar2022CFSurvey,chen2022survey,Masoumi_TWC_2020,Chien_TWC_2020,Bashar_TCOM_2021}. Some of the techniques involve different cooperation levels among the distributed units, signal compression strategies to communicate between the central and distributed units, sequential wired fronthaul, etc. A millimeter wave wireless fronthaul which provides flexible deployment but requires \gls{los} links has been explored in \cite{Zhang_JSAC_2020}. 
In \cite{shaik2021distributed}, a sequential fronthaul referred as radio-stripes has been researched where each \gls{ap} forwards the sufficient statistics to the \gls{cpu} which computes the bit \glspl{llr} to perform the data decoding. However, this suffers from fronthaul latency issues which worsen as the number of \glspl{ap} increases. 
To the best of our knowledge, none of the existing papers propose a systematic, spectrally efficient and scalable procedure that utilizes the wireless fronthaul architecture and \gls{ota} computation mechanism to combine the data in cell-free massive \gls{mimo} systems.

One of our objectives is obtain a function of the signals from the \gls{ap} to the \gls{cpu} that is sufficient to decode the information symbols. We provide an efficient and novel method of combining the data from the \glspl{ap} \gls{ota}~\cite{csahin2023survey,yang2020federated,hellstrom2022wireless,shao2022bayesian}. Such \gls{ota} schemes also find applications in collaborative machine learning or federated learning, where the local model updates sent by the distributed nodes are aggregated at a central node to obtain a global model. 
The OTA computation aims to estimate a function, $f(x_1,\ldots,x_L)\rightarrow \mathbb{C}$, of the signals transmitted by $L$ different nodes, $x_{l}$, $\forall l \in [\numAPs] \triangleq \{1.\ldots,\numAPs\}$. These functions can range from simple functions such as sum to complex functions such as geometric mean, maximum and minimum of function. This involves appropriate pre-processing and post-processing techniques at the \glspl{ap} and the \gls{cpu} to achieve the centralized system performance.

We summarize the main contributions of this paper below: 
\begin{enumerate}
    \item We provide a framework to compute the sufficient statistics \gls{ota} to decode the information symbols of the \glspl{ue} in a spectrally efficient manner in cell-free massive \gls{mimo} systems. It is important to note that the developed framework is scalable with the number of \glspl{ap}.
    \item We develop transmit precoding and two-phase power assignment schemes to combine the sufficient statistics \gls{ota} which leads to reduced communication and computation load on the wireless fronthaul.
    \item We derive the first and second order statistics of the locally available sufficient statistics i.e., the Gramian of the channel matrix and the matched filter outputs at the \glspl{ap}.
    \item We provide closed-form analytical expressions for the \gls{ls} and \gls{lmmse} estimators of the sufficient statistics.
    \item We numerically evaluate the \gls{nmse}, \gls{ser}, and the coded \gls{ber} performances of the proposed solution and benchmark against the state-of-the-art.
\end{enumerate}

\section{System Model and Preliminaries}\label{SystemModel}
We consider the \gls{ul} of a cell-free massive \gls{mimo} system with $\numUEs$ single-antenna \glspl{ue} transmitting data to $\numAPs$ \glspl{ap} with $\numAntennasPerAP$ receive antennas each. We assume that $\numAPs\numAntennasPerAP~>~\numUEs$. Each \gls{ap} transfers a linear transformation of its received signal to an $M$ receive antenna \gls{cpu} which performs the \gls{ul} data decoding. We assume that the \glspl{ap} use a different frequency band with a separate \gls{rf} chain and a dedicated wireless fronthaul link to communicate with the \gls{cpu}. We use a quasi-static block fading model for all the wireless channels in the network. 
We denote the channel between \gls{ue} $k$ and \gls{ap} $l$ by $\chan{kl}\in \complexm{N}{1}$ and the channel between \gls{ap} $l$ and the \gls{cpu} by $\chanMatCent{l} \in \complexm{M}{N}$. 

The received signal at the $l$-th \gls{ap} during the \gls{ul} data transmission phase is given by
\begin{equation}
	\mathbf{y}_l = \sqrt{\rho_{\rm ul}}\chanMat{l}\mathbf{s} + \mathbf{n}_l \in \complexm{\numAntennasPerAP}{1}
\end{equation}
where $\rho_{\rm ul}$ is the \gls{ul} \gls{snr} of each \gls{ue}
, $\chanMat{l} = \left[\chan{1l},\chan{2l},\ldots,\chan{Kl}\right]\in \complexm{N}{K}$ is the channel matrix whose columns are independent of each other and distributed as circularly symmetrically complex Gaussian $\chan{kl}\sim \CN{\mathbf{0}}{\mathbf{R}_{kl}}$, $k\in [\numUEs]$, $\mathbf{s} = [s_{1},s_{2},\ldots,s_{K}]^T \in \complexm{\numUEs}{1}$, $s_k$ is the transmit symbol of the \gls{ue} $k$, and $\mathbf{n}_l \sim \CN{\mathbf{0}}{\mathbf{I}_N}$ is the circularly symmetric additive white complex Gaussian noise at the \gls{ap} $l$ where $\mathbf{I}_N$ denotes an identity matrix of dimension $N$. To expose the concept of the \gls{ota} combining in cell-free massive \gls{mimo} systems, we assume that the \glspl{ap} have perfect \gls{csi} of the \glspl{ue}.

In a conventional system, each \gls{ap} has to send $\mathbf{y}_l$ and $\chanMat{l}$, $\ l\in [\numAPs]$, to the \gls{cpu} using orthogonal resources to decode the data, which leads to a huge fronthaul communication overhead of $(\numAPs\numAntennasPerAP + \numAPs\numAntennasPerAP\numUEs)$ complex channel uses. However, by preprocessing $\mathbf{y}_l$ and $\chanMat{l}$ at the \glspl{ap} and using efficient analog \gls{ota} combining methods in the wireless fronthaul, we can achieve the performance of a centralized decoding method with a reduced fronthaul communication overhead. One potential way to do this is to send the local sufficient statistics: the Gramian matrix $\mathbf{T}_{l}^{(1)}\triangleq \gram{\mathbf{H}_l}$ and the matched filter output $\mathbf{t}_{l}^{(2)} \triangleq \mathbf{H}_l^H\mathbf{y}_l$ rather than $\mathbf{H}_l$ and $\mathbf{y}_l$, $l\in [L]$ from the \glspl{ap} to the \gls{cpu}. This is because, any linear or non-linear multiuser detector employed at the \gls{cpu} needs only the sum of the locally obtained sufficient statistics 
\begin{equation}\label{eqn:suffstats}
	\begin{aligned}		\mathbf{T}^{(1)}=\sum_{l=1}^{L}\mathbf{T}_{l}^{(1)}
 \quad\text{and}\quad
		\mathbf{t}^{(2)}=\sum_{l=1}^{L}\mathbf{t}_{l}^{(2)}
	\end{aligned}
\end{equation}
to decode the data symbols. For example, the \gls{lmmse} detector depends only on the terms in \eqref{eqn:suffstats} as follows:
\begin{equation}\label{eqn: lmmseEst}
    \widehat{\mathbf{s}} = \sqrt{\rho_{\rm ul}}\left(\rho_{\rm ul}\mathbf{T}^{(1)} + \mathbf{I}_{K} \right)^{-1}\mathbf{t}^{(2)}.
\end{equation}
Moreover, when we utilize the wireless properties to compute \eqref{eqn:suffstats} \gls{ota}, the solution scales well with the number of \glspl{ap}.

Each \gls{ap} preprocesses and sends its locally obtained sufficient statistics to the \gls{cpu} in a non-orthogonal manner. We divide this into two phases: the $l$-th \gls{ap} transmits $\mathbf{T}_{l}^{(1)}$ and $\mathbf{t}_{l}^{(2)}$ in the first and second phases, respectively.\footnote{Refer \cite{caire2010multiuser} for more details on analog and digital feedback mechanisms.} 
We assume that $\numAntennasPerAP \geq \numAntennasCPU$ and each \gls{ap} transmits $\numAntennasCPU$ complex symbols per channel use. Further, observe that $\mathbf{T}_{l}^{(1)}$ is a Hermitian symmetric matrix and it is enough to transmit either its upper or lower triangular parts i.e., \gls{ap} $l$ transmits $\mathbf{x}_{l}^{(1)}$ given in \eqref{eqn:Phase1Data} in the next page, which is a vectorized form of the upper triangular part of $\mathbf{T}_{l}^{(1)}$. This translates to $M_1 = \left\lceil \frac{K(K+1)}{2M} \right\rceil$ transmissions in the first phase. In the second phase, $M_2 = \left\lceil \frac{\tauu K}{M} \right\rceil$ transmissions are needed, where $\tauu$ is the number of channel uses by each \gls{ue} to transmit its \gls{ul} data. 
\begin{figure*}
    \begin{equation}
		\mathbf{x}_{l}^{(1)}=  
  {\begin{bmatrix}
				\norm{\mathbf{h}_{1l}}^2 & \mathbf{h}_{1l}^H\mathbf{h}_{2l} & \cdots & \mathbf{h}_{1l}^H\mathbf{h}_{Kl} & \norm{\mathbf{h}_{2l}}^2 & \mathbf{h}_{2l}^H\mathbf{h}_{3l}& \cdots & \mathbf{h}_{2l}^H\mathbf{h}_{Kl}& \cdots &
				\norm{\mathbf{h}_{Kl}}^2
			\end{bmatrix}^T}.  
 \label{eqn:Phase1Data}
\end{equation}
\hrule \vspace{-4mm}
\end{figure*}
In the second phase, the $l$-th \gls{ap} transmits  
\begin{equation}
	\mathbf{x}_{l}^{(2)} = 
	\begin{bmatrix}
		{\mathbf{t}_{l,1}^{(2)}}^T  & {\mathbf{t}_{l,2}^{(2)}}^T  & \cdots & 
		{\mathbf{t}_{l,\tauu}^{(2)}}^T
	\end{bmatrix}^T,\label{eqn:Phase2Data}
\end{equation}
where the second index in the subscript of $\mathbf{t}_{l,t}^{(2)}$ denotes the $t$-th transmission interval of the matched filter output $\mathbf{t}_{l}^{(2)}$. We next describe the methods to combine the sufficient statistics.

\section{Over-the-Air Combining Schemes}
We propose \gls{ota} schemes to coherently combine the local sufficient statistics transmitted by the \glspl{ap} to the \gls{cpu}. 
To start with, let us denote the transmit precoder of the  $l$-th \gls{ap} by $\mathbf{W}_l\in\complexm{N}{M}$. The received signal at the \gls{cpu} is given by
\begin{align}
	\mathbf{Z}^{(i)} = \sum_{l=1}^L \sqrt{\rho_{\rm c}} \mathbf{G}_l^H\mathbf{W}_l         \bar{\mathbf{X}}_{l}^{(i)} + \mathbf{E}^{(i)}\in\complexm{M}{M_i},\label{eqn:RxSig1_CPU}
\end{align}
where $i\in\{1,2\}$ denotes the index of the two transmission phases of the \glspl{ap} to send their local sufficient statistics, $\mathbf{G}_l^H\in\complexm{M}{N}$ is the channel between the $l$-th \gls{ap} and the \gls{cpu}, $\rho_{\rm c}$ is a common power assignment scaling factor of the \glspl{ap} to satisfy the average transmit power constraint, $\mathbf{E}^{(i)}$ is the additive noise whose columns are complex Gaussian distributed with mean $\mathbf{0}_M$ and covariance $\mathbf{I}_M$. The transmit signal matrix of the $l$-th \gls{ap} denoted by $\bar{\mathbf{X}}_{l}^{(i)}$ is
\begin{equation}
	\bar{\mathbf{X}}_{l}^{(i)} = 
	\begin{bmatrix} \bar{\mathbf{x}}_{l,1}^{(i)}&\ldots&\bar{\mathbf{x}}_{l,M_i}^{(i)}
	\end{bmatrix}
	\in\complexm{M}{M_i},\quad i\in\{1,2\},\label{eqn:Xbarmat}
\end{equation}
where $\bar{\mathbf{x}}_{l,m}^{(i)}\in\complexm{M}{1}$ corresponds to the $((m-1)M+1)$-th to the $mM$-th entries of $\mathbf{x}_{l}^{(i)}$ given in \eqref{eqn:Phase1Data} and \eqref{eqn:Phase2Data}. Note that $\bar{\mathbf{x}}_{l,M_i}^{(i)}$, $i\in\{1,2\}$ are zero padded if $M$ is not an integer multiple of $\frac{K(K+1)}{2}$ and $\tauu K$ for $i=1$ and $i=2$, respectively. 

To minimize the fronthaul load, we consider the case where the \gls{cpu} does not need to have any knowledge of the \gls{csi} of the \glspl{ap}. The \glspl{ap} obtain their respective \gls{csi} using downlink training from the \gls{cpu}. Moreover, for ease of exposition and also because \glspl{ap} and the \gls{cpu} are stationary, we assume perfect \gls{csi} at the \glspl{ap}. To obtain the sum of the local sufficient statistics, each \gls{ap} first offsets the effect of $\mathbf{G}_l$, $l\in[L]$. In this paper, we assume that all the channels between the \glspl{ap} and the \gls{cpu} have full column rank. In future generation wireless systems, the \glspl{ap} can be low-cost nodes deployed by the network planners based on the requirements and therefore the channels to the \gls{cpu} undergo rich scattering which justifies the full column rank assumption. We will deal with other channel models in the journal version of this paper. We set the number of antennas at the \gls{cpu} less than or equal to that of the \glspl{ap} ($M\leq N$) and use a local \gls{zf} precoder at each \gls{ap} as $\mathbf{W}_l = \mathbf{G}_l\left(\mathbf{G}_l^H \mathbf{G}_l\right)^{-1}\in\complexm{N}{M}$, $l\in [L]$. As these channels are relatively time invariant, each \gls{ap} computes the transmit precoder once and reuses it till the channels change. 

We recall that our goal is to compute $\sum_{l=1}^{L}\bar{\mathbf{X}}_{l}^{(i)}$, and \gls{zf} precoding only cancels the effects of the channels between the \glspl{ap} and the \gls{cpu}. However, any unequal power scaling at the \glspl{ap} will result in residual scaling factors which are impossible to be removed \gls{ota}. To circumvent this power scaling problem, we propose an average transmit power assignment strategy at the \glspl{ap} to communicate with the \gls{cpu}. The total energy expended by the $l$-th \gls{ap} during the $M_1$ and $M_2$ symbol intervals is:
\begin{align}
	\Omega_l^{(i)} &= \rho_{\rm c}^{(i)}\,\sum_{t=1}^{M_i} \expectL{\norm{\mathbf{W}_l\bar{\mathbf{x}}_{l,t}^{(i)}}^2},\nonumber\\
	&=\rho_{\rm c}^{(i)}\, \trace{\left(\mathbf{I}_{M_i}\otimes\expectL{\mathbf{W}_l^H\mathbf{W}_l}\right)\expectL{{\mathbf{x}}_{l}^{(i)} {{\mathbf{x}}_{l}^{(i)^H}}} },\label{eqn:ForPowerConstraint}
\end{align}
where $i\in\{1,2\}$ and $\otimes$ denotes the matrix Kronecker product operator.\footnote{To obtain \eqref{eqn:ForPowerConstraint}, we assume that $\frac{K(K+1)}{2}$ and $\tauu K$ are multiples of $M$. However, we can handle the general case with a minor modification.} We use two separate scaling mechanisms for transmitting the Gramian and the matched filter outputs from the \glspl{ap} to the \gls{cpu} to accommodate for the differences in their dynamic ranges. Note that the matched filter outputs have $\rho_{\rm ul}$ embedded in them which changes their dynamic ranges compared to that of the Gramian matrices. Therefore, it is imperative to employ this two-phase power assignment strategy to obtain good system performance.

The values of $\Omega_l^{(1)}$ and $\Omega_l^{(2)}$ can be computed using the corresponding mean and covariance matrices of the sufficient statistics which we will discuss subsequently. 
Therefore, the average transmit power of the $l$-th \gls{ap} to transmit the Gramian matrices and the matched filter outputs of one coherence interval is
\begin{align}
	P_{l}^{(i)} = \frac{\Omega_l^{(i)}}{M_i},\qquad i\in\{1,2\}.\label{eqn:AvgTransPower}
\end{align}

Now, let us denote the maximum average power constraint by $P_{\rm max}$ for all the \glspl{ap}. We propose a low overhead feedback mechanism to compute the common power scaling factors $\rho_{\rm c}^{(1)}$ and $\rho_{\rm c}^{(2)}$ at the \gls{cpu} (to assist in the coherent addition of the sufficient statistics at the \gls{cpu}) and broadcast it to the \glspl{ap} via a common control channel to satisfy the average transmit power constraint. 

For $i\in\{1,2\}$, each \gls{ap} sets $\rho_{\rm c}^{(i)}$ equal to $1$, computes the scalar value in \eqref{eqn:AvgTransPower} using the locally available statistics and conveys it to the \gls{cpu} using a dedicated control channel. 
This can be done after the pilot transmissions from the \gls{cpu} to the \glspl{ap} occasionally. Upon receiving $\{P_{1}^{(i)}, \ldots, P_{L}^{(i)}\}$, $i\in\{1,2\}$, the \gls{cpu} obtains the indices of the \glspl{ap} which violate the constraint. Let us include these indices in the sets $\mathcal{V}^{(i)} = \{i_1^{(i)}, \ldots, i_{L'^{(i)}}^{(i)}\}\subseteq\{1,\ldots,L\}$, where $L'^{(i)}$, $i\in\{1,2\}$, is the number of \glspl{ap} in it. Then the scaling factors are computed as
\begin{align}
    \rho_{\rm c}^{(i)} = \frac{P_{\rm max}}{\max_{i\in\mathcal{V}} P_{i}^{(i)}},\qquad i\in\{1,2\}.\label{eqn:computescalefactor1}
\end{align}

This ensures that every \gls{ap} satisfies its average transmit power constraint and is necessary to add the sufficient statistics coherently at the \gls{cpu}.

Substituting $\mathbf{W}_l = \mathbf{G}_l\left(\mathbf{G}_l^H \mathbf{G}_l\right)^{-1}$ in to the received signal \eqref{eqn:RxSig1_CPU}, the \gls{cpu} receives the following signal
\begin{equation}
	\mathbf{Z}^{(i)} \triangleq \begin{bmatrix}
		\mathbf{z}_1^{(i)}&\ldots&\mathbf{z}_{M_i}^{(i)}
	\end{bmatrix}= \sqrt{\rho_{\rm c}^{(i)}}\sum_{l=1}^{L}\bar{\mathbf{X}}_{l}^{(i)} + \mathbf{E}^{(i)},
\end{equation}
where $i\in\{1,2\}$, $\{\mathbf{z}_1^{(i)},\ldots,\mathbf{z}_{M_i}^{(i)}\}$ are the columns of $\mathbf{Z}^{(i)}$ and $\bar{\mathbf{X}}_{l}^{(i)}$ is given in \eqref{eqn:Xbarmat}. 

Now we provide two estimators to obtain the sum of the local sufficient statistics at the \gls{cpu}: Bayesian and classical estimators. To do that, we first derive the mean and covariance of the sufficient statistics below.

\subsection{Derivation of the Statistics of the Sufficient Statistics}\label{sec:SuffStatsDeriv}

First, note that the channels between the \gls{ap} and the \glspl{ue} are independent of each other which means that $\{\mathbf{x}_{l}^{(1)}\}$, $l\in [L]$ are statistically independent. Therefore, the mean $\boldsymbol{\mu}^{(1)}$ and the covariance matrix $\mathbf{C}^{(1)}$ of $\mathbf{x}^{(1)}=\sum_{l=1}^L\mathbf{x}_{l}^{(1)}$ are given by
\begin{align}
    \boldsymbol{\mu}^{(1)} = \sum_{l=1}^L\boldsymbol{\mu}_{l}^{(1)},\qquad
    \mathbf{C}^{(1)} =\sum_{l=1}^L\mathbf{C}_{l}^{(1)},\label{eqn:mu1C1stats}
\end{align}
where $\boldsymbol{\mu}_{l}^{(1)}$ and $\mathbf{C}_{l}^{(1)}$ are the mean and covariance matrix of $\mathbf{x}_{l}^{(1)}$, $l\in [\numAPs]$. To compute $\boldsymbol{\mu}_{l}^{(1)}$ and $\mathbf{C}_{l}^{(1)}$ when the channels between the \glspl{ue} and the \glspl{ap} undergo correlated Rayleigh fading, we need the statistics of the sum of chi-squared and product of complex Gaussian random variables. We compute them in closed form here. The nonzero entries of $\boldsymbol{\mu}_{l}^{(1)}$ and $\mathbf{C}_{l}^{(1)}$ are governed by the following indexing equations: For a  given $j\in[K]$ and $j' \in \{j,\ldots,K\}$, the non-zero $n$-th entry of $\boldsymbol{\mu}_{l}^{(1)}$ and non-zero $(n',n')$-th diagonal entry of $\mathbf{C}_{l}^{(1)}$ are at the indices $n = (K-0.5j)(j-1) + j$ and $n' = (K-0.5j)(j-1)+j'$ which are given by
\begin{align}
	\boldsymbol{\mu}_{l}^{(1)}[n] =\trace{\mathbf{R}_{jl}},\quad 
	\mathbf{C}_{l}^{(1)}[n',n'] = \trace{\mathbf{R}_{jl}\mathbf{R}_{j'l}}\ \label{eqn:C1l}
\end{align}
For $i=2$, to compute the mean and covariance of each column in \eqref{eqn:Xbarmat}, we focus on the term $\sum_{l=1}^L\mathbf{t}_{l,t}^{(2)}$ i.e., the matched filter output at time $t\in[\tau_u]$. For $t\neq t'$, $\mathbf{t}_{l,t}^{(2)}$ and $\mathbf{t}_{l,t'}^{(2)}$ are uncorrelated and for $l\neq l'$, $\mathbf{t}_{l,t}^{(2)}$ and $\mathbf{t}_{l',t}^{(2)}$  are correlated. Therefore, to compute the mean and covariance of $\sum_{l=1}^L\mathbf{t}_{l,t}^{(2)}$, we need the mean $\boldsymbol{\mu}_{l,t}^{(2)}$, the covariance matrix $\mathbf{C}_{l,t}^{(2)}$ of $\mathbf{t}_{l,t}^{(2)}$ and also the cross-covariance matrix $\mathbf{C}_{ll',t}^{(2)}$ between $\mathbf{t}_{l,t}^{(2)}$ and $\mathbf{t}_{l',t}^{(2)}$. We can compute the mean to be zero because $\boldsymbol{\mu}_{l,t}^{(2)}=\mathbf{0}$ for any $l$ and $t$. The covariance $\mathbf{C}_{t}^{(2)}$ of $\sum_{l=1}^L \mathbf{t}_{l,t}^{(2)}$ is given by
\begin{equation}
    \mathbf{C}_{t}^{(2)} = \sum_{l=1}^L\mathbf{C}_{l,t}^{(2)}   + \sum_{l=1}^L\sum_{\substack{l'=1,l'\neq l}}^L\mathbf{C}_{ll',t}^{(2)}.\label{eqn:C2l1}
\end{equation}
For correlated Rayleigh fading, we derive the closed form expressions as
\begin{align}
	\mathbf{C}_{l,t}^{(2)} &= \rho_{\rm ul}\expectL{\left(\mathbf{H}_l^H\mathbf{H}_l\right)^2}  + \expectL{\left(\mathbf{H}_l^H\mathbf{H}_l\right)},\label{eqn:C2l}\\
 \mathbf{C}_{ll',t}^{(2)} &= \rho_{\rm ul}\expectL{\left(\mathbf{H}_l^H\mathbf{H}_l\right)}\expectL{\left(\mathbf{H}_{l'}^H\mathbf{H}_{l'}\right)}.\label{eqn:C2lm}
\end{align}
where $\expect{\left(\mathbf{H}_l^H\mathbf{H}_l\right)} = {\rm diag}\left(\trace{\mathbf{R}_{1l}},\ldots,\trace{\mathbf{R}_{Kl}}\right)$, and $\expect{\left(\mathbf{H}_l^H\mathbf{H}_l\right)^2}$ is a diagonal matrix whose $k$-th diagonal entry is $\trace{\mathbf{R}_{kl}}^2 + \trace{\mathbf{R}_{kl}\sum_{k' = 1}^{K}\mathbf{R}_{k'l}}$. Finally, $\expect{\mathbf{W}_l^H\mathbf{W}_l}$ can be evaluated numerically. With the mean, covariance and cross covariance matrices of the sum of the sufficient statistics, we present the estimators to obtain them below.

\subsection{\Gls{lmmse} and \Gls{ls} Estimators}
With a prior on the sufficient statistics, the \gls{cpu} considers a \gls{lmmse} estimator of $\sum_{l=1}^{L}\bar{\mathbf{X}}_{l}^{(i)}$, $i\in\{1,2\}$. For convenience, let us denote the mean and covariance of $\sum_{l=1}^{\numAPs}\bar{\mathbf{x}}_{l,m}^{(i)}$. $m\in[M_i],\ i\in\{1,2\}$ by $\bar{\boldsymbol{\mu}}_m^{(i)}$ and $\bar{\mathbf{C}}_m^{(i)}$, respectively. Note that $\bar{\boldsymbol{\mu}}_m^{(i)}$ and $\bar{\mathbf{C}}_m^{(i)}$ can be obtained by selecting the appropriate entries from the mean and covariance matrices derived in \eqref{eqn:mu1C1stats} and \eqref{eqn:C2l1}.
Then, the \gls{lmmse} estimate of $\sum_{l=1}^L \bar{\mathbf{x}}_{l,m}^{(i)}$, $i\in\{1,2\}$ (denoted by $\widehat{\bar{\mathbf{x}}}_m^{(i)}$), for $m\in[M_i],\ i\in\{1,2\}$ is
\begin{align}\label{eqn:LMMSEestimatSuffStats}
	\widehat{\bar{\mathbf{x}}}_m^{(i)} =\bar{\boldsymbol{\mu}}_m^{(i)} + \sqrt{\rho_{\rm c}^{(i)}}\bar{\mathbf{C}}_m^{(i)}&\left({\rho_{\rm c}^{(i)}}\bar{\mathbf{C}}_m^{(i)} + \mathbf{I}_{M}\right)^{-1}\nonumber\\
    &\times\left(\mathbf{z}_m^{(i)} - \sqrt{\rho_{\rm c}^{(i)}}\bar{\boldsymbol{\mu}}_m^{(i)}\right).
\end{align}

In the case when the \gls{cpu} does not have any prior information of the sufficient statistics, the \gls{cpu} implements the \gls{mvu} estimator
\begin{equation}\label{eqn:LSestimatSuffStats}
	\widehat{\bar{\mathbf{x}}}_m^{(i)} =\left({\rho_{\rm c}^{(i)}}\right)^{-\frac12}\mathbf{z}_m^{(i)},
\end{equation}
which is also efficient \cite{kay1993fundamentals}.  In this case, the \gls{mvu} estimator is also the \gls{ls} estimator. Finally, the \gls{cpu} 
computes the Gramian matrix $\widehat{\gram{\mathbf{H}}}$ and the matched filter output $\widehat{\mathbf{H}^H\mathbf{y}}$ in the two phases using the estimated sum of the sufficient statistics.

\section{Data Detection}
We use the sum of the sufficient statistics obtained during the two phases to detect the information symbols of the \glspl{ue}. Let us denote $\mathbf{y} = \begin{bmatrix}\mathbf{y}_1^H &\ldots &\mathbf{y}_\numAPs^H\end{bmatrix}^H \in \mathbb{C}^{\numAPs\numAntennasPerAP}$, $\chanMat{} = [\chanMat{l}^H,\ldots,\chanMat{\numAPs}^H]^H \in \complexm{\numAPs\numAntennasPerAP}{\numUEs}$. We use the following data detectors for our numerical evaluation. Note that our developed solution is equally applicable to any data detector.

\subsection{Linear Detectors}
We present the centralized \gls{lmmse} and \gls{ls} estimates of the data symbols here:
\begin{equation}\label{eqn:approxLMMSE_Est}
    \begin{aligned}
    \widehat{\mathbf{s}}_{\text{\gls{lmmse}}} &=  \sqrt{\rho_{\rm ul}}\left(\rho_{\rm ul}\gram{\mathbf{H}} + \mathbf{I}_{K} \right)^{-1}\mathbf{H}^H\mathbf{y}\\
        &\approx  \sqrt{\rho_{\rm ul}}\left(\rho_{\rm ul}\widehat{\gram{\mathbf{H}}} + \mathbf{I}_{K} \right)^{-1}\widehat{\mathbf{H}^H\mathbf{y}}.
    \end{aligned}    
\end{equation}
and 
\begin{equation}
    \widehat{\mathbf{s}}_{\text{LS}} = \rho_{\rm ul}^{-1/2}\left(\widehat{\gram{\mathbf{H}}}\right)^{-1}\widehat{\mathbf{H}^H\mathbf{y}},
\end{equation}
respectively.

\subsection{\Gls{map} Detector}
The \gls{map} detector outputs
\begin{equation*}
   \begin{aligned}
        \widehat{\mathbf{s}}_{\text{MAP}} &=  \argmax_{\mathbf{s}\in \mathcal{S}} \norm{\mathbf{y} - \sqrt{\rho_{\rm ul}}\mathbf{H}\mathbf{s}}= \argmax_{\mathbf{s}\in \mathcal{S}} \norm{\overline{\mathbf{y}} - \sqrt{\rho_{\rm ul}}\overline{\mathbf{H}}\mathbf{s}}\\
         &\approx  \argmax_{\mathbf{s}\in \mathcal{S}} \norm{\widehat{\overline{\mathbf{y}}} -\sqrt{\rho_{\rm ul}}\widehat{ \overline{\mathbf{H}}}\mathbf{s}},
   \end{aligned}
\end{equation*}
where ${\overline{\mathbf{H}}} = \left({\gram{\mathbf{H}}}\right)^{\frac{1}{2}}$, ${\overline{\mathbf{y}} }= \left({\gram{\mathbf{H}}}\right)^{-\frac{1}{2}}$$\mathbf{H}^H\mathbf{y}$, \\$\widehat{\overline{\mathbf{H}}} = \left(\widehat{\gram{\mathbf{H}}}\right)^{\frac{1}{2}}$ and $\widehat{\overline{\mathbf{y}} }= \left(\widehat{\gram{\mathbf{H}}}\right)^{-\frac{1}{2}}\widehat{\mathbf{H}^H\mathbf{y}}$ and $\mathcal{S}$ is the constellation set.

\subsection{Soft-output Detection}
We can also use soft detection methods to compute the bit \glspl{llr} as:
\begin{equation}\label{eqn: llrSig}
    \mathcal{L}\left(b_i\rvert \widehat{\mathbf{H}^H\mathbf{y}}, \widehat{\gram{\mathbf{H}}}\right) \approx \ln\left(\frac{\sum_{\mathbf{s}:b_i(\mathbf{s})=1}e^{-\norm{\widehat{\overline{\mathbf{y}}} - \sqrt{\rho_{\rm ul}}\widehat{\overline{\mathbf{H}}}\mathbf{s}}^2}}{\sum_{\mathbf{s}:b_i(\mathbf{s})=0}e^{-\norm{\widehat{\overline{\mathbf{y}}} - \sqrt{\rho_{\rm ul}}\widehat{\overline{\mathbf{H}}}\mathbf{s}}^2}}\right)
\end{equation}
where $\mathcal{L}(b_i) = \ln ({\rm Pr}(b_i=1)/{\rm Pr}(b_i=0))$, and the notation $\mathbf{s}:b_i(\mathbf{s})=\alpha$ means the set of all vectors $\mathbf{s}$ for which the $i$-th bit is $\alpha$ i.e., $b_i(\mathbf{s})=\alpha$. After computing the \glspl{llr}, we can input them to a channel decoder in case of a coded communication system.

\section{Numerical Results}\label{numericalResults}
In this section, we evaluate the performance of the proposed \gls{ota} framework using the following simulation setup: The locations of the \glspl{ue} are uniformly distributed in a square area of $200~{\rm m}\times 200~{\rm m}$ whereas the locations of \glspl{ap} and the \gls{cpu} are fixed around the center of the square. 
We further position the antennas of the \glspl{ap} and the \gls{cpu} at a height of  $5~{\rm m}$ above the \glspl{ue}. We consider a 3GPP urban microcell channel propagation model with a carrier frequency of $2$~GHz~\cite[Table~B.1.2.1-1]{LTE2010b}. The large-scale fading coefficient is determined by: $\beta_{kl} \triangleq {\trace{\mathbf{R}_{kl}}}/{N} = -30.5 - 36.7\log_{10}({\rm d}_{kl}/{1{\rm m}})$,
where ${\rm d}_{kl}$ represents the distance between the \gls{ap} $l$ and the \gls{ue} $k$. The noise power spectral density, bandwidth and the noise figure are set to $-174~{\rm dBm/Hz}$, $1~{\rm MHz}$ and $5~{\rm dB}$, respectively. We set the values of $\numAPs$, $\numUEs$, $\numAntennasPerAP$, and $M$ to $16$, $8$, $5$, and $4$, respectively. The \glspl{ap} and the \gls{cpu} use uniform linear array antennas with half-wavelength antenna spacing. The spatial correlation matrices, $\mathbf{R}_{kl},\ k\in [\numUEs],l\in[\numAPs]$ are modeled as uncorrelated channels. With spatially uncorrelated channels between \glspl{ap} and the \gls{cpu}, we can compute $\expect{\mathbf{W}_l^H\mathbf{W}_l}$ using closed form expressions~\cite{tague1994expectations}. We denote $\rho_{\rm ul} = \frac{P_{\rm ul}\beta_{\rm avg}}{BN_0}$, where $P_{\rm ul}$ is the transmit power of the \glspl{ue}, $B$ is the bandwidth, $N_0$ is the additive white noise power spectral density, and $\beta_{\rm avg}$ is the average path loss of a \gls{ue}.

\begin{figure}[htbp]
	\centering
	\includegraphics{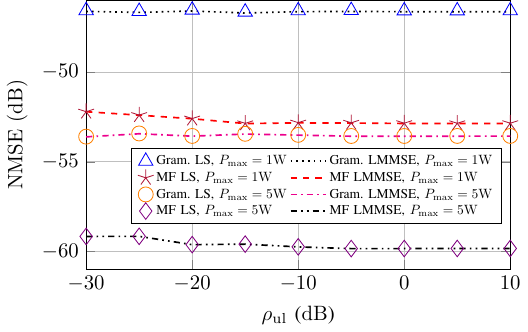}
	\caption{\gls{nmse} of the estimated sufficient statistics as a function of $\rho_{\rm ul}$.}\vspace{-3mm}
	\label{plotNMSE_SuffStats}
\end{figure}

In Fig.~\ref{plotNMSE_SuffStats}, we plot the \gls{nmse} in ${\rm dB}$ of the estimated Gramian and matched filter outputs (defined as $\frac{\expect{\norm{\mathbf{x} - \widehat{\mathbf{x}}}^2}}{\expect{\norm{\mathbf{x}}}^2}$, where $\widehat{\mathbf{x}}$ is the estimate of a random vector $\mathbf{x}$ and the expecation operator performs an empirical average of at least $10^5$ Monte carlo runs.) 
as a function of $\rho_{\rm ul}$. We compare the \gls{lmmse} and \gls{ls} estimators in \eqref{eqn:LMMSEestimatSuffStats} and \eqref{eqn:LSestimatSuffStats} when $P_{\rm max}$ is set to $1$~W and $5$~W. 
We observe that the performance difference between the \gls{ls} and the \gls{lmmse} estimators is negligible and both of them achieve an \gls{nmse} of less than $-45$~dB even when $P_{\rm max}$ is set to only $1$~W. Note that the \gls{nmse} of the Gramian estimate is a constant for any value of $\rho_{\rm ul}$. The Gramian matrices transmitted by the \glspl{ap} during the first phase are independent of the transmit \glspl{snr} of the \glspl{ue} which results in a constant \gls{nmse}. However, as $P_{\rm max}$ increases, the \gls{nmse} of the Gramian matrix decreases. 
{On the other hand, the matched filter outputs transmitted by the \glspl{ap} and the power scaling factor $\rho_{\rm c}^{(2)}$ are dependent on $\rho_{\rm ul}$ which impacts their \gls{mse} performance. However, when we normalize the \gls{mse} with the expected power of the matched filter outputs, it results in almost a constant line at higher transmit \glspl{snr} as shown in the Fig.~\ref{plotNMSE_SuffStats}.}

\begin{figure}[htbp]
	\centering
	\includegraphics{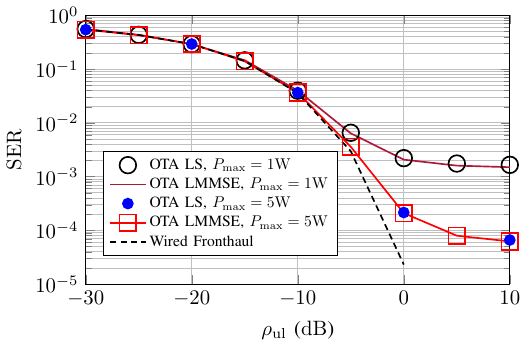}
	\caption{\gls{ser} of the detected data symbols as a function of $\rho_{\rm ul}$.}
	\label{plotSER_data}\vspace{-5mm}
\end{figure}

In Fig.~\ref{plotSER_data}, we show the \gls{ser} performance of the \gls{lmmse} data detector as a function of $\rho_{\rm ul}$ when the \glspl{ue} transmit 4-\gls{qam} modulated symbols to the \glspl{ap}. The sufficient statistics are obtained using \gls{ls} and \gls{lmmse} estimators. An initial symbol estimate is obtained using the \gls{lmmse} detector in \eqref{eqn:approxLMMSE_Est} followed by a nearest-neighbor approximation to detect the data symbols. We observe that the \gls{ser} performance with the sufficient statistics estimated through \gls{ls} and \gls{lmmse} estimators almost matches that of a cell-free massive \gls{mimo} system with centralized decoding upto around $-15~{\rm dB}$ with $P_{\rm max} = 5~{\rm W}$. An interesting observation is that the \gls{ser} obtained by the \gls{ota} methods reaches an error floor beyond a particular value of $\rho_{\rm ul}$. Moreover, as $P_{\rm max}$ increases, the value of $\rho_{\rm ul}$ at which the error floor happens also increases. This is because, as $\rho_{\rm ul}$ increases, the scaling factor $\rho_{\rm c}^{(2)}$ to satisfy the average transmit power constraint during the transmission of the matched filter outputs from the \glspl{ap} decreases. This leads to a saturation effect in the receive \gls{snr} at the \gls{cpu} resulting in an error floor. Further, as $P_{\rm max}$ increases, the saturation effect of the receive \gls{snr} at the \gls{cpu} occurs at a higher value of $\rho_{\rm ul}$. 

\begin{figure}[htbp]
	\centering
	\includegraphics{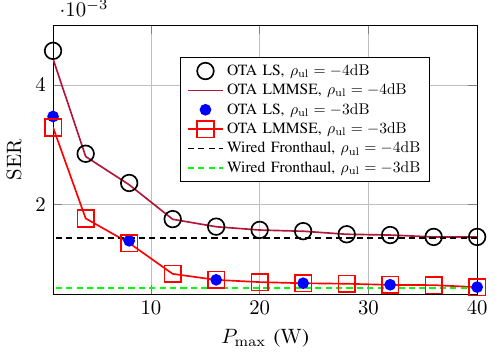}
	\caption{\gls{ser} of the detected data symbols as a function of $P_{\rm max}$.}
	\label{plotSER_data2}\vspace{-5mm}
\end{figure}

We plot the \gls{ser} as a function of $P_{\rm max}$ in the Fig.~\ref{plotSER_data2} when $\rho_{\rm ul}$ is set to $\{-4,-3\}~{\rm dB}$. We observe that as $P_{\rm max}$ increases, the reliability of the wireless links between the \glspl{ap} and the \gls{cpu} improves which leads to reduction in the \gls{ser} for both the values of $\rho_{\rm ul}$. Moreover, the performance difference between the \gls{ls}, \gls{lmmse} estimators and a wired fronthaul is negligible beyond a $P_{\rm max}$ of around $30$~dB. This also confirms our earlier observation that increasing $P_{\rm max}$ alleviates the error floor issue seen in the Fig.~\ref{plotSER_data}. This shows that there is a performance-to-power consumption trade-off and appropriate selection of the system and transmit parameters can be done based on the system design requirements. 

\begin{figure}[htbp]
	\centering
	\includegraphics{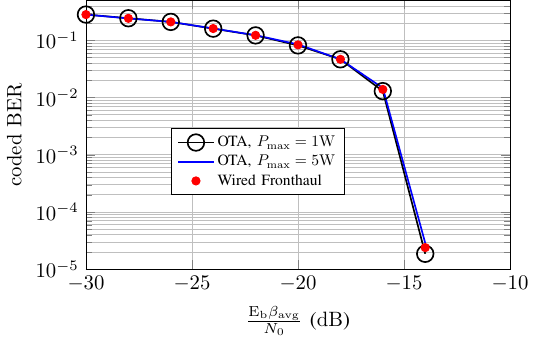}
	\caption{Coded \gls{ber} as a function of \gls{snr}. Sphere decoder is used to obtain the posterior bit \glspl{llr}.}
	\label{plotCodedBER}\vspace{-2mm}
\end{figure}

In Fig.~\ref{plotCodedBER}, we plot the coded \gls{ber} as a function of $\frac{{\rm E_b}\beta_{\rm avg}}{N_0}$ when $P_{\rm max}$ is set to $1$~W and $5$~W. We employ an LDPC error correcting code of rate ${\rm R_c} = 1/2$ and length $1944$ from the IEEE 802.11-2020 \gls{wlan} standard~\cite{standardIEEE802_11_2020}. We set $K$, $N$ and $M$ to $4$, $5$, and $3$, respectively. The symbol energy ${\rm E_s}$ is related to the bit energy ${\rm E_b}$ as ${\rm E_s} = {\rm E_b}\log_2(|\mathcal{S}|){\rm R_c}$, where $|\mathcal{S}|$ is the cardinality of the constellation set $\mathcal{S}$. We have used a max-log approximation of \eqref{eqn: llrSig} to compute the posterior bit \glspl{llr} 
and adopted the sphere decoding algorithm to obtain them efficiently. We have used the \gls{lmmse} estimator to obtain the sufficient statistics and benchmarked the coded \gls{ber} performance with a centralized wired fronthaul based system. We observe that channel coding not only reduces the performance gap between the wireless and wired fronthaul based systems, but also assists in mitigating the error floor issue seen in the Fig.~\ref{plotSER_data}. Note that there will be an error floor at much higher \glspl{snr} which may be beyond the region of our interest. 

\section{Conclusions}\label{conclu}
In this paper, we presented a novel scalable framework to combine the locally obtained sufficient statistics at the \glspl{ap} \gls{ota} in cell-free massive \gls{mimo} systems. 
We presented the \gls{ls} and \gls{lmmse} estimators to obtain the sum of the sufficient statistics using the closed-form expressions of the statistics of the sufficient statistics for the Gaussian channel case. We also developed the transmit precoding and two-phase power assignment mechanisms to coherently combine the sufficient statistics \gls{ota}. 
We empirically observed that the \gls{nmse}, \gls{ser} and coded \gls{ber} performances of the developed \gls{ota} framework closely matches that of a wired fronthaul based cell-free massive MIMO system. Our numerical results show that the developed \gls{ota} framework is a feasible and scalable approach to reduce the communication and computational overhead associated with a wired fronthaul based cell-free massive \gls{mimo} system. 
\bibliographystyle{IEEEtran}
\bibliography{IEEEabrv,reff}
\end{document}